# Explanation of the Phase Diagram of High-Temperature Superconductors in Terms of the Model of Negative-U Centers Superconductivity

K.D. Tsendin[1], B.P. Popov[2], D.V. Denisov[1]

[1] *A. F. Ioffe Physico-Technical Institute, Russian Academy of Sciences, 194021, Saint-Petersburg, Russia*

[2] *Saint-Petersburg State Polytechnical University, 195251, Saint-Petersburg, Russia*

ABSTRACT

It is demonstrated qualitatively how a unified explanation of the wide variety of different regions in a typical phase diagram of HTSC can be given in terms of the model of negative-U centers superconductivity. Both the existence of four regions (pseudogap, superconducting, non Fermi-liquid and Fermi-liquid), in the phase diagram and all the transitions between these regions were explained by assuming that negative-U centers exist in HTSC and qualitatively taking into account their thermodynamic and direct quantum-mechanical interaction with ordinary electrons from the valence band.

It is found that the principal parameter determining the properties of HTSC in the model of negative-U centers is the relative concentration of electrons belonging to negative-U centers. The negative-U centers naturally determine both the superconducting properties of HTSC systems and the entire phase diagram in the normal state.

INTRODUCTION

It is known that high-temperature superconductors exhibit a number of unusual properties in both normal and superconducting states. In this connection, the phase diagram of high-temperature superconductors (HTSC) contains a large number of characteristic regions.

Figure 1 shows schematically a typical phase diagram of a HTSC in the coordinates "temperature-doping level" [1]. In the general case, the relative concentration of holes per unit cell of $CuO_2$ is commonly plotted along the abscissa axis. Along the same axis, we plotted the real content of the dopant (oxygen) for one of the best-studied systems, $Y_1Ba_2Cu_3O_x$, because we are going to use the parameters of this system by the way of an example.

For x falling within the range ~6.0-6.4 samples do not undergo a superconducting transition (region AF delimited by curve $T_N$ ) and, at low temperature, are Hubbard antiferromagnetic (AF) insulators with a composition-dependent Néel temperature $T_N(x)$. As the temperature increases within the range $T_p > T > T_N$, the samples pass into the so-called pseudogap (PG) state. The region occupied by this state (PG region) extends as far as x = 6.9-7.0, i.e., includes the compositions that can be transferred to the superconducting state at a temperature $T_c(x)$. In this case, the pseudogap state exists in the temperature interval $T_p > T > T_c$. The characteristic dome-shaped curve $T_c(x)$ delimits the region of existence of the superconducting phase (SC region). Above this region lies the region of the normal phase, delimited by the lines $T_p(x)$ and $T_L(x)$. For the latter region, numerous deviations from the Fermi-liquid behavior are observed (NFL region). Finally, the FL region lies at a high content of oxygen, where the superconductivity virtually disappears and HTSC behave as ordinary Fermi-liquid metals.

The properties of samples in the AF region are commonly easily understood. It is generally accepted that the antiferromagnetic state of the samples is due to spin states of $Cu^{2+}$ ions in $CuO_2$ planes. Therefore, we do not consider this region in what follows.

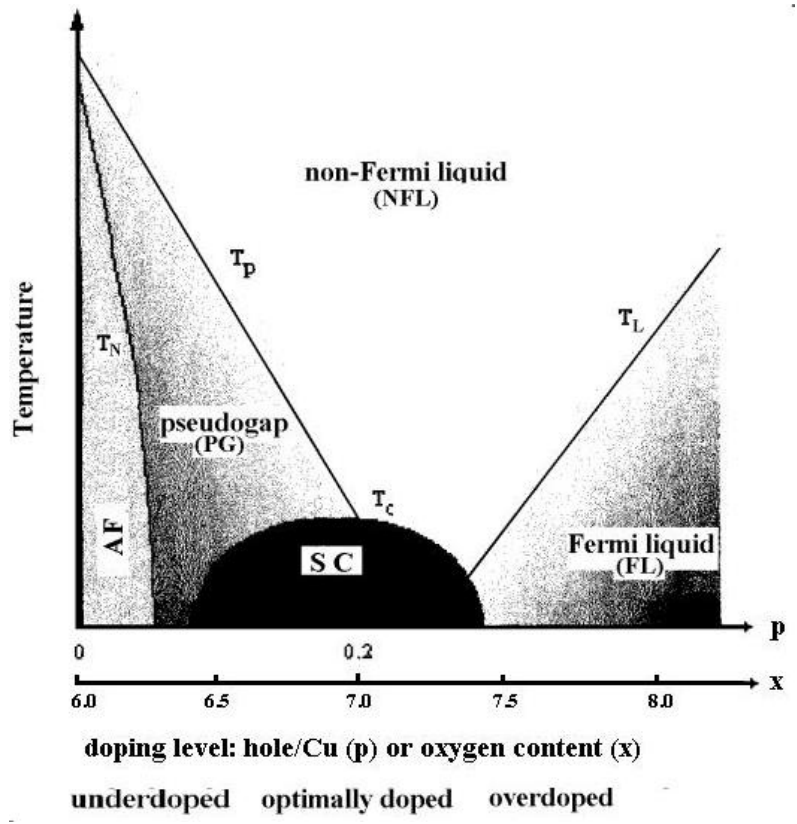

*Figure 1. A typical phase diagram of a HTSC in the coordinates "temperature-doping level" [1]. P is the relative concentration of holes per unit cell of $CuO_2$. X is the real content of the oxygen for $Y_1Ba_2Cu_3O_x$.*

The present communication is aimed to explain properties of the other four regions. At present, there exist numerous approaches that account both for the properties of these regions and for transitions between them. For example, recent reviews [2, 3] summarized the results obtained on applying a somewhat modified, but still the classical approach based on the BCS theory. True, mostly the optimal-doping region was considered in this case. The difference between the properties of HTSC and those of the classical superconductors [4, 5] has led to a wide use of non-phonon mechanisms of electron pairing, which were considered in detail in the reviews [4-8]. However, it should be stated that there is no unified explanation of the whole phase diagram both in terms of the classical approach [2, 3] and from the standpoint on non-phonon mechanisms.

In this communication, we are going to demonstrate that a unified explanation of the entire phase diagram can be given in terms of the model of negative-U centers [9, 10].

The concept of negative-U centers was first put forward by Anderson in 1975 for describing some properties of chalcogenide glassy semiconductors [11] and was further developed in [12, 13]. It was assumed that centers with a specific property exist in the atomic lattice of a material. The strong electron-lattice interaction results in that the binding energy of two electrons exceeds that of their Coulomb repulsion. Such an effect is also observed under normal conditions, i.e., electrons coupled into a pair exist even at room and higher temperatures. For superconductivity to appear, it suffices that the pairs could move and the nondegenerate Bose gas would become degenerate, i.e., all the moving pairs would pass into the coherent state. The possibility that superconductivity can appear in such a system was noted before the BCS theory was developed. This idea was first suggested by Ogg in 1946 [14]; later, such a possibility was analyzed in detail by Shafroth in 1955 [15]. Models based on this concept became widely accepted after Bednorz and Muller discovered the first high-temperature superconductors based on cuprates [16]. At the International conference on fundamental aspects of superconductivity (Russia, Zvenigorod, 2004), several possible mechanisms of formation of electron pairs at temperatures considerably exceeding $T_c$ were reported [17-20]. A detailed analysis of reports indicating that negative-U centers exist in HTSC can be found in the review [21].

In this communication, we disregard mechanisms of formation of negative-U centers, and only assume that they are present in HTSC. Below, we first remind the basic concept of the model of negative-U centers [9, 10]. Then we consider in detail its statistical properties and explain on its basis the properties of all of the above-mentioned four regions of the phase diagram of HTSC.

## MODEL OF NEGATIVE-U CENTERS SUPERCONDUCTIVITY

The model of negative-U centers superconductivity (NUCS model) [9, 10] is based on the results of theoretical studies [22-25], in which superconducting properties of pairs moving over the system of negative-U centers were considered.

In the one-electron energy band diagram, the energies of the first and second ionization of an isolated negative-U center are $E_1$ and $E_2$, respectively. These energies are commonly denoted by arrows that connect the $D^-$ and $D^+$ levels of the negative-U center and the conduction band (Fig. 2a). According to [22-25], the system of interacting negative-U centers can be described by a Hubbard Hamiltonian with a negative effective correlation energy, whose absolute value is equal to the difference of the energies of the $D^-$ and $D^+$ levels.

$$H = -U \cdot \sum n_{i\uparrow} \cdot n_{i\downarrow} + \sum t_{ij} \cdot a_{i\sigma}^+ \cdot a_{j\sigma} \qquad (1)$$

where $n_{i\sigma} = a_{i\sigma}^+ \cdot a_{i\sigma}$ are occupation numbers; $a_{i\sigma}^+$ and $a_{i\sigma}$ are operators of creation and annihilation of an electron with a spin $\sigma$ at site i; and $t_{ij}$, matrix element of a one-electron transition between the nearest localization centers (negative-U centers). $U > 0$ and it is assumed that $t_{ij} = t \ll U$. The negative values of -U lead to attraction of electrons with opposite spins at a site. It is assumed that the binding energy of this coupling exceeds the ordinary Coulomb correlation energy, i.e., the resultant interaction -U in (1) is negative. It is assumed in the model that, at a sufficiently high concentration of negative-U centers, the $D^-$ and $D^+$ levels are broadened into the respective $W^-$ and $W^+$ bands with a total width $2W = 2zI$ for a simple cubic lattice of negative-U centers [22] (Fig. 2b). Because U considerably exceeds t, the contribution of real one-particle transitions in the system of negative-U centers can be neglected and $W^-$ and $W^+$ are bands of transfer of strongly coupled pairs (bosons), with an effective matrix element of transition of a pair equal to $I = 2t^2/U$. At a temperature $T_c$, Hamiltonian (1) gives rise to a superconducting correlation between pairs, i.e., to Bose-condensation in $W^-$ and $W^+$ bands. According to [22], $T_c$ is given by

$$T_c = W \cdot (1-2\nu)/\ln(\nu^{-1}-1) \qquad (2)$$

where $\nu$ is the relative concentration of pairs, equal to n/2D (n is the concentration of electrons in the system of negative-U centers, and D, concentration of negative-U centers).

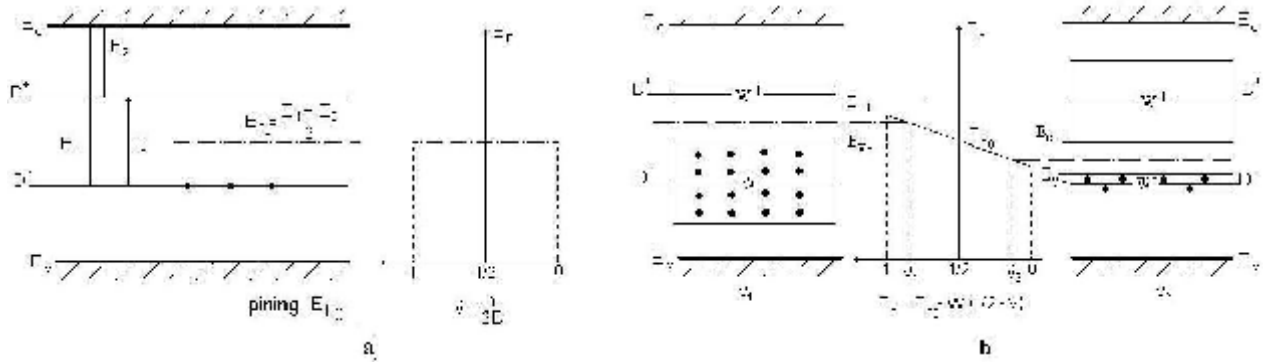

*Figure 2. Dependencies of Fermi level $E_F$ on $\nu$ for t=0 (a) and t≠0 (b).*

In [9, 10, 19, 26, 27], a model of negative-U centers superconductivity (NUCS model) was formulated and elaborated on the basis of the results obtained. The main points of the model are as follows.

(1) The system contains negative-U centers with a binding energy U considerably exceeding the one-electron matrix elements responsible for transitions of electrons between the centers: t<<U. The concentration of negative-U centers is sufficiently high, so that, because of the non-vanishing value of t, they constitute a transport system over which preliminarily formed pairs can move. The superconducting phase transition occurs due to Bose condensation of these pairs.

(2) In addition to electron pairs belonging to the system of negative-U centers, there exist "ordinary" band electrons weakly interacting with the lattice. As a first approximation, the statistical and direct quantum-mechanical interaction of "ordinary" electrons and electrons belonging to the system of negative-U centers is considered only qualitatively.

The statistical interaction means redistribution of whole electron concentration between two groups: "ordinary" band electrons and electrons belonging to the system of negative-U centers.

The direct quantum-mechanical interaction means the changing of electron wave functions due to coincidence or proximity of one-particle energies of electrons from two different groups.

As regards the basic concept, our NUCS model is similar to models of preformed pairs [14, 15, 17-19, 21] or bipolaron models [20, 28, 29]. However, a very important difference from these latter consists is that our model considers two groups of electrons and takes into account the interactions between these groups.

The model of negative-U centers superconductivity (NUCS model) has made it possible to explain several important experimental facts.

(a) If we assume, for estimation purposes, that negative-U centers are situated at sites of a simple cubic lattice and substitute t = 50 meV [30] and U = 1.8 eV [10, 17] into (2), we obtain for z = 6 and ν = 1/2 a value of $T_c$ equal to the maximum critical temperature (~90 K) for the $Y_1Ba_2Cu_3O_x$ system.

(b) According to formula (2), the dependence of $T_c$ on ν is dome-shaped, with a maximum at ν = 1/2. Therefore, as shown in [10], formula (2) quite naturally explains the dome-shaped dependence of the temperature of the superconducting transition on the doping level, observed experimentally in quite a number of HTSC systems.

(c) We demonstrated in [19, 26, 27] that the pseudogap-related features of conductivity in underdoped samples and the effect of additional conductivity in overdoped samples can be understood on the assumption that the mid-distance between the W- and W+ bands lies slightly higher than the top of the valence band for underdoped samples and inside the valence band, close to its edge, in the case of overdoped samples. Thus, it was shown that specific features of the conductivity of under- and overdoped HTSC systems can be considered from a common standpoint and accounted for by the relative positions of the Fermi level and the top of the valence band.

In our previous communications [19, 26, 27], the relative positions of the Fermi level and of the top of the valence band, necessary for explaining experimental data, were postulated.

The aim of this study was to demonstrate that the change in the relative positions of $E_F$ and $E_v$ is determined by the properties of the NUCS model itself. For this purpose, we consider the statistical properties of the system of negative-U centers and then construct the phase diagram of HTSC on the basis of the results obtained.

Thus, it will be shown that two points of the model, formulated above, provide an insight into all specific features of HTSC, i.e., furnish an understanding of the entire phase diagram of HTSC without making any additional assumptions.

STATISTICAL PROPERTIES OF THE SYSTEM OF NEGATIVE-U CENTERS

One of the most important statistical properties of negative-U centers is the pinning of the Fermi level by them centers at zero temperature [11, 31]. Figure 2a shows the pinning of the Fermi level (for the case of t=0) between the levels of charged states of negative-U centers, $D^-$ and $D^+$. In the left-hand part of the figure, $E_{1,2}$ are the energies of the first and second ionizations of negative-U centers and U is the absolute value of the binding energy of pairs. For convenience, we consider the case when the $D^-$ and $D^+$ levels lie in the energy gap. When the electron concentration n in the system changes, the concentrations of $D^-$ and $D^+$ centers change too, whereas the Fermi level remains in an invariable position $E_{F0}$ for all relative electron pair concentrations $\nu = n/2D$ in the range $0 < \nu < 1$ (plot in the right-hand part of Fig.2a). The situation resembles that in an intrinsic semiconductor, with the occupied $D^-$ level acting as a completely filled "valence band," and the Fermi level $E_{F0}$, with energy equal to the average energy of ionization per electron, $(E_1 + E_2)/2$, lying in the middle between this "valence band" and the empty "conduction band," whose role is played by the $D^+$ level. Full circles in the figure represent pairs of coupled electrons.

Let us demonstrate now how taking into account the finite value of t in terms of the NUCS model quite naturally leads to a Fermi level that depends (at zero temperature) on the doping level. As t ≠ 0, the $D^-$ and $D^+$ levels are broadened in this case into $W^-$ and $W^+$ bands (Fig. 2b). In the one-electron diagram in Fig. 2b, these bands have an unusual appearance: their width depends on the degree ν of their filling. To understand and use this fact let us disregard pair breaking processes and denote the total width of the bands $W^-$ and $W^+$ by 2W:

$$W^- + W^+ = 2W \qquad (3)$$

Then the spacing between the closest levels in the $W^-$ and $W^+$ bands is given by

$$\Delta E = 2W/D \qquad (4)$$

because only a single state for a pair exists at each negative-U center and the entire band of pair transport has a width 2W. Using the parameter ΔE, the widths of the filled band $W^-$ and the totally empty band $W^+$ can be written as follows:

$$W^- = (n/2)\ \Delta E = (n/2)\ 2W/D = 2\nu W \qquad (5a)$$
$$W^+ = (D-n/2)\ \Delta E = (D-n/2)\ 2W/D = 2(1-\nu)W \qquad (5b)$$

because, owing to the pinning of the Fermi level, the lower band $W^-$ must be always occupied by pairs with a concentration n/2 and the upper band $W^+$ must be always empty. As for the pinning of the Fermi level, it must now occur at mid-distance between the top ($E_{v-}$) of the $W^-$ band and bottom ($E_{c+}$) of the $W^+$ band. Let us take as zero energy the position of the $D^+$ level, then the $D^-$ level has an energy (-U). Assuming that the broadening of levels into bands is symmetric with respect to the positions of the $D^-$ and $D^+$ levels, we obtain that the $E_{v-}$ and $E_{c+}$ levels are shifted toward each other by half widths of the $W^-$ and $W^+$ bands, respectively.

$$E_{v-}=-U+W^-/2 = -U+\nu W \qquad (6a)$$

$$E_{c+}=-(1-\nu)W \qquad (6b)$$

Then the Fermi level lying between these band edges is given by

$$E_F=(E_{v-}+E_{c+})/2=-U/2-W(1/2-\nu)=E_{F0}-W(1/2-\nu) \qquad (7)$$

Its dependence on the degree of filling, $\nu$, is shown in Fig. 2b. As $\nu$ decreases from 1 to 0, the Fermi energy linearly decreases and the Fermi level passes the position $E_{F0}$ at half-filling ($\nu=0.5$). For clarity, Fig. 2b shows band diagrams for degrees of filling larger ($\nu_1$) and smaller ($\nu_2$) than 0.5 to the left and to the right of the $E_F(\nu)$ dependence.

EXPLANATION OF THE PHASE DIAGRAM OF HTSC

Let us now apply the results obtained to explanation of the phase diagram of HTSC.

**REGION OF PSEUDOGAP FEATURES (PG)**

In our opinion, the mid-distance between the $W^-$ and $W^+$ bands lies near the top of the true valence band, $E_v$, [19, 29, 30]. Let us first assume for simplicity that, in the case of half-filling, the Fermi level coincides with the top of the true valence band: $E_{F0}=E_v$ (Fig. 3a). To make Fig. 3 less cumbersome, the $W^+$ bands are not shown.

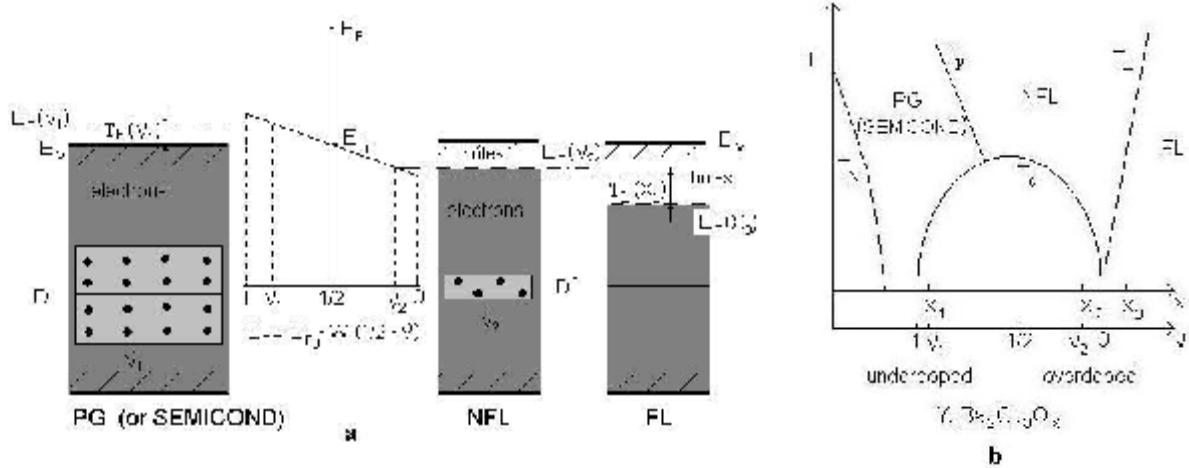

*Figure 3. Energy band diagrams for PG, NFL and FL regions (a). And whole phase diagram of $Y_1Ba_2Cu_3O_x$ in the NUCS model (b).*

Then the mid-distance between the $W^-$ and $W^+$ bands, i.e., the Fermi level at zero temperature, lies above $E_v$ at $1/2 < \nu < 1$ and, at low temperatures, the sample resistance has a typical semiconducting temperature dependence with an activation energy equal to $(E_F - E_v)$. This situation is illustrated for a particular value of $\nu$ ($\nu_1$) in Fig. 3a (PG case). The valence band fully occupied by "ordinary" electrons is represented in the figure by the shaded region. As in Fig. 2, the $W^-$ band is represented by a region with full circles designating pairs of electrons that belong to negative-U centers. As temperature increases, the activation law of hole creation in the valence band virtually ceases to be obeyed at a temperature approximately equal to $T_p=(E_F-E_v)$ and the temperature dependence of the resistance is mainly determined by mobility. Thus, the points of the $T_p(x)$ curve are, in our interpretation, temperatures of crossover from the semiconducting temperature dependence of the resistance in the PG region to a metallic dependence in the NFL region. Therefore, the PG (pseudogap) region in the phase diagram in Fig. 3b could be named the region of semiconductors (semicond.), for which the conductivity is determined by holes from the valence band and the position of the Fermi level, which lies within the energy gap at zero temperature determined by negative-U centers. However, the metallic conductivity in the NFL region must also exhibit specific features of non-Fermi-liquid nature in the general case, because the direct quantum-mechanical mixing of states belonging to negative-U centers and states of "ordinary" electrons from the valence band remains strong at $T>T_p$.

As temperature is lowered, we pass from the PG region to the region of superconductivity (SC, Fig. 3b). At $T = T_c$ the sample becomes superconducting, which is due to Bose condensation of hole pairs belonging to the $W^+$ band. According to our previous publication [10], the dependence $T_c(x)$ is described in this case by the left-hand part of the dome-shaped dependence $T_c(x)$ shown in Fig. 3b. In this figure, the quantity ν, which is a variable that is the most important in this study, is plotted along with the oxygen content x, instead of the relative hole concentration p in Fig. 1. The phenomenological relationship between ν and x (ν = 7.4 - x) was qualitatively substantiated in [10].

In Fig. 3a, the upper boundary of the valence band, $E_v$, is the mobility edge for this band, and the tail of the density of localized states is not shown. Thus, underdoped HTSC from the PG region are, in our interpretation, Fermi-glasses with a Fermi level lying on the background of localized states whose density is responsible for the signal intensity in photoemission experiments.

**REGION OF NON-FERMI-LIQUID BEHAVIOR (NFL) IN THE NORMAL STATE**

Let us first consider this region at values of ν, approximately falling within the range 1/2 <ν < 1. In this case, we pass from the PG region to the region NFL of non-Fermi-liquid behavior at temperatures exceeding $T_p(x)$ (see Fig. 3b). In our model, the main reason for the non-Fermi-liquid behavior in the normal state is the strong resonant direct quantum-mechanical (not thermodynamic) interaction of electrons belonging to negative-U centers with electrons from the valence band. In the ν range under consideration, this interaction becomes important when the temperature exceeds the energy difference $(E_F-E_v)=T_p$.

If we leave aside the superconducting transition, then, at 0 < ν < 1/2, the Fermi level at zero temperature enters the valence band, i.e. lies below $E_v$ (Fig. 3a, NFL case). Outwardly, the band diagram of the normal state of HTSC looks like a diagram of a classical metal with a valence band filled with electrons (shaded region) up to $E_F$. However, we have, in fact, not an ordinary metal because at $0 < \nu < 1/2$ in the normal state, the Fermi level is still pinned at mid-distance between the top ($E_{v-}$) of the $W^-$ band and the bottom ($E_{c+}$) of the $W^+$ band. Electrons that left band states lying above the Fermi level contribute to the occupancies of the bands of negative-U centers and do not determine directly the position of $E_F$ at these $\nu$. In the $\nu$ range under consideration, the superconductivity is due to Bose condensation of electrons from the $W^-$ band. As the temperature increases to $T > T_c(x)$, there occurs a transition to the NFL region, because the quantum-mechanical interaction of electrons from the valence band with electrons belonging to negative-U centers is of the strongest, resonant nature. Figure 3a (NFL case) shows the situation described for a particular value of $\nu = \nu_2$.

## REGION OF FERMI-LIQUID (FL) BEHAVIOR IN THE NORMAL STATE

Let us now consider the region of doping level, lying to the right of the value $\nu = 0$. In this case, the $W^-$ band disappears because all the negative-U centers are already only in the $D^+$ states. This means that the Fermi level at last moves away from the mid-distance between the edges of the $W^-$ and $W^+$ bands and, as x changes further (for the negative-U centers $\nu$ is always zero, and the relation $\nu=7.4-x$ ceases to be valid), its position will reflect the decreasing number of electrons in the valence band. In other words, we finally obtain the situation of a classical Fermi-liquid metal, the FL region. However, such a situation is only preserved at low temperatures: $T \ll [(E_{c+}(\nu=0) - E_{v-}(\nu=0))/2 - E_F] = T_L$, when thermal electrons close to the Fermi level do not "feel" the existence of negative-U centers. At temperatures of about $T_L$ and higher, an effective quantum-mechanical interaction of ordinary electrons from the valence band with electrons belonging to negative-U centers sets in, and we again have a non-Fermi-liquid behavior, i.e., pass from the FL region to the NFL region. As the condition $\nu = 0$ is satisfied for the entire FL region, all values in Fig. 3a (FL case) are given for a particular $x_3$.

CONCLUSION

It was demonstrated qualitatively how a unified explanation of the wide variety of different regions in a typical phase diagram of HTSC can be given in terms of the model of negative-U centers superconductivity [9, 10, 19, 26, 27]. Both the existence of four regions, PG, SC, NFL, and FL, in the phase diagram and all the transitions between these regions were explained by assuming that negative-U centers exist in HTSC [17-21] and taking into account their thermodynamic and direct quantum-mechanical interaction with ordinary electrons from the valence band.

It was found that the principal parameter determining the properties of HTSC in the model of negative-U centers is the relative concentration $\nu$ of electrons belonging to negative-U centers. Just this quantity predetermines the characteristic dome-shaped dependence of the temperature of the superconducting phase transition on composition. This same quantity governs the variation of the relative positions of levels related to negative-U centers and to ordinary electrons from the valence band, which makes it possible to describe the presence of all regions of the normal state in the phase diagram.

In our previous studies [19, 26, 27], the relative positions of the Fermi level and of the top of the valence band, necessary for interpretation of experimental data, was postulated. In this study, it was shown that the change in the relative positions of $E_F$ and $E_v$ is determined by the properties of the model of negative-U centers itself. Thus, it was shown that two points of the model: presence of negative-U centers and their interaction with ordinary electrons from the valence band, make it possible to understand, without making any additional assumptions, all the specific features of HTSC, i.e., the entire phase diagram of HTSC.

It should be emphasized that, in the NUCS model under consideration the negative-U centers naturally determine both the superconducting properties of HTSC systems and the entire phase diagram in the normal state.

REFERENCES

1. Batlog B.,.Varma C.M., *Phys World*, February, 33 (2000).
2. Maximov E.G., *Physics-Uspekhi* **170**, 1033 (2000).
3. Maximov E.G., *Physics-Uspekhi* **174**, 1026 (2004).
4. Ginzburg V.L., *Physics-Uspekhi* **170**, 619 (2000).
5. Kopaev Yu.V., *Physics-Uspekhi* **159**, 567 (1989).
6. Iz'umov Yu.A., *Physics-Uspekhi* **169**, 225 (1999).
7. Sadovskiy M.V., *Physics-Uspekhi* **171**, 539 (2001).
8. Bel'avskiy V.I., Kopaev Yu.V., *Physics-Uspekhi* **174**, 457 (2004)
9. Popov B. P., Tsendin K. D., *Tech. Phys. Lett.* **24**, 265 (1998).
10. Tsendin K.D., Popov B.P., *Supercond. Sci. Technol.* **12**, 255 (1999).
11. Anderson P.W., Phys. *Rev. Lett.* **34,** 953 (1975).
12. Street R.F., Mott W.F., *Phys. Lett.* **35**, 1293 (1975).
13. Kastner M., Adler D., Fritzsche H., *Phys. Rev. Lett.* **37**, 1504 (1976).
14. Ogg R., *Phys. Rev.* **69**, 243 (1946).
15. Shafroth M.R., *Phys. Rev. B* **100**, 460 (1955).
16. Bednorz J. G., Muller K.A.Z., *Phys. B* **64**, 189 (1986).
17. Andreev A.F. *JETP Letters* **79**, 100 (2004)
18. Micen K.V., Ivanenko O.M., *Zh. Eksp. Teor. Fiz.* **118**, 666 (2000). Micen K.V., Ivanenko O.M., Abstr. of 1st Intern. Confer. On Fund. Problems of HTSC (in Russian), Moscow-Zvenigorod, 48 (2004).
19. Tsendin K.D., Denisov D.V., Popov B.P., Abstr. of 1st Intern. Confer. On Fund. Problems of HTSC (in Russian), Moscow-Zvenigorod, 201 (2004).
20. Alexandrov A.S. Abstr. of 1st Intern. Confer. On Fund. Problems of HTSC (in Russian), Moscow-Zvenigorod, 4 (2004).
21. Wilson J., *J. Phys.: Cond. Mat.* **13**, R945 (2001).
22. Kulik I.O., Pedan A.G., *Sov. Phys. JETP* **52** 742 (1980).
23. Kulik I.O., Pedan A.G., *Sov. J. Low Temp. Phys.* **8,** 118 (1982)
24. Kulik I.O., Pedan A.G., *Sov. J. Low Temp. Phys.* **9**(3), 127 (1983).

25. Bulayevsky L.N., Sobyanin A.A., Khomsky D.I, *Sov. Phys.-JETP* **60** 865 (1984).
26. K.D. Tsendin, D.V. Denisov, *Supercond. Sci. Technol.*, **16**, 80 (2003).
27. Tsendin K.D., Denisov D.V., Popov B.P., *JETP Letters* **80**, 246 (2004).
28. Alexandrov A. Ranninger J., *Phys. Rev. B* **23,** 1796; **24,** 1164 (1981).
29. Alexandrov A.S., Ranninger J., Robaszkiewicz, *Phys. Rev. B* **33** 4526 (1986).
30. Eremin M.V., Lavizina O.V., *Sov. Phys.-JETP* **111**, 144 (1997).
31. Adler D. Yoffa E.F., *Phys. Rev. Lett*. **36**, 1197 (1976).